\documentstyle[12pt,epsf,aps]{revtex}

\newcommand{\be}{\begin{equation}}
\newcommand{\ee}{\end{equation}}
\newcommand{\bea}{\begin{eqnarray}}
\newcommand{\eea}{\end{eqnarray}}

\begin{document}


\title
{
The Character of the Principal Series of Representations
of the Real Unimodular Group
 \vspace*{5mm}
}
\author
{
 Debabrata Basu\footnote{email: dbasu@phy.iitkgp.ernet.in}
}
\address
{
 Department of Physics, 
 Indian Institute of Technology, \\
 Kharagpur - 721302, West Bengal, INDIA.
}

\maketitle


\newpage
\begin{abstract}
\begin{center}
{\large \bf Abstract}
\end{center}
The character of the principal series of representations 
of $SL(n,R)$ is evaluated by using Gel'fand and Naimark's
definition of character.
This representation is realized in the space of functions
defined on the right coset space of $SL(n,R)$ with respect
to the subgroup of real triangular matrices.
This form of the representations considerably simplifies
the problem of determination of the integral kernel of
the group ring which is fundamental in the Gel'fand-Naimark
theory of character. An important feature of the principal
series of representations is that the `elliptic' elements
of $SL(n,R)$ do not contribute to its character. 
\end{abstract}

\newpage

\section{Introduction}
{\label{intro}}

One of the principal problems in the theory of representations
of the real unimodular group of arbitrary order is the determination of
the irreducible unitary representations of the discrete series. In their
attempt to determine all the unitary irreducible representations
(UIR's) of this family of groups 
Gel'fand and Graev\cite{gelf_graev} regarded the
group $ G = SL(n,R)$ as the real form of 
$ G^c = SL(n,C)$. The discrete series of UIR's which exists
only for even $n$ appears in this construction
in an especially interesting way.
These are distinguished by the fact that they are defined in the
space of purely analytic functions. For $n=2$ the space $W$
of the complex matrices on which the group acts splits
into two transitive subspaces $W^+$ and $W^-$ consisting of 
functions analytic in the upper and lower half-plane respectively.
These are the positive and negative discrete series of the UIR's
of $SL(2,R)$ (or $SU(1,1)$). Although the Gel'fand - Graev
theory is perhaps the only theory capable of explaining
the occurrence of the discrete series of representations
of $SL(n,R)$ the complete resolution of this problem still
remains open.

For the principal series of representations however, the above 
involved realization of the representation can be dispensed with, 
and the character of the representation can be determined 
by a simple extension\cite{gelf_naimark} of 
Gel'fand and Naimark's method as
outlined in their seminal paper\cite{gelf_naimark_paper} 
on the Lorentz group.
In two previous papers\cite{basu_1996,basu_1999} this problem
was completely solved for $n=2$. The character of the discrete
series of representations was determined by using the 
Hilbert space method of Bargmann\cite{bargmann199a}
and Segal\cite{segal} while those of the principal 
and exceptional series were determined by a simple extension
of the method of ref.\cite{gelf_naimark_paper}.
The main advantage of this method is that the entire analysis 
can be carried out within the canonical framework
of Bargmann\cite{bargmann199b}.
It is the object of this paper to extend this method 
for the evaluation of the character of the principal
series of representations of the real unimodular group
of arbitrary order.

First of all it is necessary to realize the representations
of the principal series of $SL(n,R)$ in a form suitable
for the computation of the character. In what follows
the representations of the principal series will be realized
in the form of operators in the space of the functions defined
on the right coset space of $SL(n,R)$ with respect to the 
subgroup of real triangular matrices. This form of the 
representation is particularly suited for the determination
of the integral kernel of the group ring which is 
fundamental in the Gel'fand~-~Naimark theory of character.
The general outline of the remaining steps of the calculation
is similar to ref.\cite{basu_1996}. An important feature
of the principal series of representations is that the
`elliptic' elements of $SL(n,R)$ i.e. the elements
of $SL(n,R)$ corresponding to complex eigenvalues do not
contribute to its character. 


\section{Some subgroups of $SL(\lowercase{n},R)$ and the invariant
measure on them}

In this analysis the triangular and diagonal subgroups 
of the real unimodular group will play a very important 
role. The representations will be constructed in the 
space of functions defined on the right coset spaces
of the group with respect to the subgroup of real 
triangular matrices.

\subsection{The subgroup $K$ and the invariant measure 
on it}

$K$ is the group of all triangular matrices of the form
\bea
k &=&
\left(
\begin{array}{cc@{~~~}c@{~~~}c}
k_{11} & k_{12} & \cdots & k_{1n} \\
0      & k_{22} & \cdots & k_{2n} \\
\vdots & \vdots & \ddots & \vdots \\
0      & 0      & \cdots & k_{nn} 
\end{array}
\right)
\eea
subject to the condition
\bea
\mbox{det}~ k 
&=& k_{11} k_{22} \ldots k_{nn} = 1
\eea
The above matrix has the following properties: 
$k_{pq} = 0$ for $p > q$, $k_{pq} \neq 0$
for $p \leq q$.
To determine the left and right invariant measures on $K$
we choose as parameters in $K$ the coordinates
$k_{pq}$, $p < q$ and $k_{pp}$ 
($p = 2,3, \ldots, n$).
We note that out of the $n$ diagonal elements 
$k_{11}, k_{22}, \ldots, k_{nn}$
only $n - 1$ diagonal elements can be regarded as 
independent because of the unimodularity condition.

We start from the left~-~invariant differential
\bea
d\omega &=& k^{-1} dk
\eea
where $dk$ is the matrix of the elements
$dk_{pq}$. 
Since the diagonal elements of the matrix 
$k \in K$ are connected by
\bea
\frac{dk_{11}}{k_{11}}
+ \frac{dk_{22}}{k_{22}}
+ \cdots
+ \frac{dk_{nn}}{k_{nn}}
&=& 0
\eea
all the elements 
$dk_{pq}$
are not independent arbitray increments.
Hence out of the $n$ diagonal elements only
$n - 1$ are independent which we choose as,
$dk_{22}, dk_{33}, \ldots, dk_{nn}$.
The off diagonal elements are all arbitrary.

Proceeding in the usual way and denoting
$k^{-1}$ by $Q$ so that
$Q_{ij} = K_{ji}$, $K_{ji}$
being the cofactor of $k_{ji}$
we obtain the left invariant measure on $K$,
\bea
d \mu_l (k)
&=&
\mid \mbox{det}~ D \mid
~\prod_{p < q} dk_{pq}
~\prod_{p=2}^{n} dk_{pp}
\eea
Here $D$ is a triangular matrix whose determinant is
given by the product of its diagonal elements
\bea
\mbox{det}~ D
&=&
Q_{11}^{n-1} ~Q_{22}^{n-1} ~Q_{33}^{n-2}
~Q_{44}^{n-3} ~\cdots ~Q_{n-1,n-1}^{2} ~Q_{nn}
\nonumber \\
&=&
k_{33} ~k_{44}^2 ~\cdots ~k_{nn}^{n-2}
\eea
Hence
\bea
d \mu_l (k)
&=&
\mid k_{33} \mid
\mid k_{44} \mid^2
\ldots
\mid k_{nn} \mid^{n-2}
~\prod_{p < q} dk_{pq}
~\prod_{p=2}^{n} dk_{pp}
\eea
For the calculation of the right invariant measure 
it is convenient to make a different choice of basis
i.e a different choice of the independent parameters
of the subgroup $K$. Of course after carrying out the 
calculations we shall transform it back to the old set.
We now arrange the independent elements of
$k \in K$ in the following sequence
\bea
k_{11}~;~ k_{12}, k_{22}~;~ k_{13}, k_{23}, k_{33}~;~ 
\ldots~;~ k_{1n}, k_{2n}, \ldots, k_{n-1,n}
\hspace{10mm} (k_{nn}~~\mbox{ommited})
\nonumber
\eea
The elements of the right invariant differential 
\bea
d\omega = dk ~ k^{-1}
\eea
are arranged in the same sequence
\bea
d\omega_{11}~;~ d\omega_{12}, d\omega_{22}~;~ 
d\omega_{13}, d\omega_{23}, d\omega_{33}~;~
\ldots~;~ d\omega_{1n}, d\omega_{2n}, \ldots, d\omega_{n-1,n}
\nonumber
\eea
Hence
\bea
d \mu_r (k)
&=&
\mid \mbox{det}~ D \mid
~\prod_{p \neq q, ~p < q} dk_{pq}
~\prod_{p=1}^{n-1} dk_{pp}
\eea
where $D$ is a triangular matrix so that its determinant is
the product of its diagonal elements. Thus
\bea
d \mu_r (k)
&=&
\mid k_{11} \mid^{-1}
\mid k_{22} \mid^{-2}
\mid k_{33} \mid^{-3}
\ldots
\mid k_{n-1,n-1} \mid^{-(n-1)}
\mid k_{nn} \mid^{-(n-1)}
~\prod_{p < q} dk_{pq}
~\prod_{p=1}^{n-1} dk_{pp}
\eea
The above measure has been calculated with 
$k_{11}, \ldots, k_{n-1,n}$ as the independent elements.
We now transform it back to the old basis in which
$k_{11}$ is not independent but $k_{nn}$ is.
Thus
\bea
dk_{11}
~=~
\left| \frac{\partial k_{11}}{\partial k_{nn}} \right|
dk_{nn}
~=~
\left| \frac{k_{11}}{k_{nn}} \right|
dk_{nn}
\eea
Hence
\bea
d \mu_r (k)
&=&
\mid k_{22} \mid^{-2}
\mid k_{33} \mid^{-3}
\ldots
\mid k_{nn} \mid^{-n}
~\prod_{p < q} dk_{pq}
~\prod_{p=2}^{n} dk_{pp}
\eea
Let us now introduce the Radon~-~Nikodym derivative
\bea
\beta(k) &~=~& \frac{d \mu_l (k)}{d \mu_r (k)}
\nonumber \\
&~=~&
\mid k_{22} \mid^{2}~
\mid k_{33} \mid^{4}~
\ldots
\mid k_{nn} \mid^{2n-2}
\label{eqn13}
\eea
From the definition of $\beta(k)$ it follows that 
\bea
\beta (k_1 k_2) 
&~=~& \beta (k_1)~ \beta (k_2)
\eea

\subsection{The subgroup $H$}

We denote by $H$ the set of all matrices
$h = ~\parallel h_{pq} \parallel$
satisfying the condition 
$h_{pq} = 0$ for $q>p$.
Thus $h$ is a triangular matrix with the upper triangle zero.
Clearly $H$ is a subgroup of $SL(n,R)$.
It is also clear that $H$ has properties analogous to 
those of $K$. These properties can be derived from those of $K$
in the following way. Let us denote by $\tilde{g}$ the tanspose
of the matrix $g$. It is evident that this operation
carries $K$ into $H$ and $H$ into $K$.
As a consequence the left shift in $K$ is equivalent to the 
right shift in $H$. It, therefore, follows that the left invariant
measure in $K$ coincides with the right invariant measure
in $H$ and the right invariant measure in $K$ coincides with the 
left invariant measure in $H$:
\bea
d \mu_r (h)
&=&
\mid h_{33} \mid
\mid h_{44} \mid^2
\ldots
\mid h_{nn} \mid^{n-2}
~\prod_{p > q} dh_{pq}
~\prod_{p=2}^{n} dh_{pp}
\eea
\bea
d \mu_l (h)
&=&
\mid h_{22} \mid^{-2}
\mid h_{33} \mid^{-3}
\ldots
\mid h_{nn} \mid^{-n}
~\prod_{p > q} dh_{pq}
~\prod_{p=2}^{n} dh_{pp}
\eea

\subsection{The subgroup $X$}

Let us denote by $X$ the subgroup of matrices 
$x = ~\parallel x_{pq} \parallel \in H$ 
all diagonal elements of which are equal to $1$.
\bea
x_{pq} &=& 0 \hspace{10mm} \mbox{for} \hspace{4mm}p<q
\nonumber \\
x_{pp} &=& 1
\eea
We choose as parameters determining $x \in X$ the 
variables $x_{pq}, p>q$.
The left and right invariant measure on $X$ are given by
\bea
d \mu_r (x) ~=~ d \mu_l (x) ~=~ \prod_{p>q} dx_{pq}
\eea

\subsection{The subgroup Z}

We denote by $Z$ the subgroup of matrices
$\zeta = ~\parallel \zeta_{pq} \parallel \in K$
all diagonal elements of which are equal to $1$.
\bea
\zeta_{pq} &=& 0 \hspace{10mm} \mbox{for} \hspace{4mm}p>q
\nonumber \\
\zeta_{pp} &=& 1
\eea
As before the left shift in $X$ is equivalent to the right
shift in $Z$ and as a consequence
\bea
d \mu_r (\zeta) ~=~ d \mu_l (\zeta) ~=~ \prod_{p<q} d \zeta_{pq}
\eea

\subsection{The subgroup $D$}

We denote by $D$ the subgroup of real diagonal matrices
\bea
\delta
&=&
\left(
\begin{array}{c@{~~~}c@{~~~}c@{~~~}c@{~~~}c}
\delta_1 & 0       & 0      & \cdots & 0 \\
0        & \delta_2 & 0      & \cdots & 0 \\
\vdots   & \vdots  & \vdots & \ddots & \vdots \\
0        & 0       & 0      & \cdots & \delta_n
\end{array}
\right)
\eea
satisfying 
\bea
\delta_1~ \delta_2~ \ldots~ \delta_n = 1
\eea
the left and right nvariant measure on $D$ is given by
\bea
d \mu (\delta)
&=&
\frac{d \delta_2~ d \delta_3~ \ldots~ d \delta_n}
     {\mid \delta_2 \mid
      \mid \delta_3 \mid \ldots
      \mid \delta_n \mid}
\eea

\subsection{Some relations among
$G = SL(n,R), H, K, X, Z, D$}

To study the cosets of the real unimodular 
group with respect to the subgroups mentioned above
it is useful to have all the elements of $SL(n,R)$,
with some exceptions, represented in the form of products
of these subgroups.

Let us first consider representations of the elements of the group
$K$. Since $Z$ and $D$ are subgroups of $K$ each product
of the form $\zeta \delta$ and $\delta \zeta $
where $\zeta \in Z$ and $\delta \in D$ is an element
of $K$. Conversely each $k \in K$ may be represented
in a unique way in the form
\bea
k ~=~ \delta~ \zeta ~=~ \zeta'~ \delta
\eea
where
\bea
k_{pp} = \delta_p
\hspace{4mm}, \hspace{10mm}
\zeta_{pq} = \frac{k_{pq}}{k_{pp}}
\hspace{4mm}, \hspace{10mm}
\zeta_{pq}' = \frac{k_{pq}}{k_{qq}}
\eea
We can derive analogous representations for the subgroup
$H$:
\bea
h ~=~ \delta~ x ~=~ x'~ \delta
\label{eqn22a}
\eea
where $\delta \in D$, $x \in X$ and
\bea
h_{pp} = \delta_p
\hspace{4mm}, \hspace{10mm}
x_{pq} = \frac{h_{pq}}{h_{pp}}
\hspace{4mm}, \hspace{10mm}
x_{pq}' = \frac{h_{pq}}{h_{qq}}
\label{eqn22b}
\eea
Let us now consider arbitrary elements of $SL(n,R)$.
We show that each element $g \in SL(n,R)$, with some exceptions,
can be represented in the form
\bea
g ~=~ \zeta~ h \hspace{4mm}; \hspace{10mm}
\zeta \in Z \hspace{3mm}, \hspace{3mm} h \in H
\eea
Only those elements are exceptional for which one (or more)
of the minors
\bea
g_m &=& \mbox{det}~ G_m
\eea
where
\bea
G_m
&=&
\left(
\begin{array}{c@{~~~}c@{~~~}c@{~~~}c}
g_{mm} & g_{m,m+1} & \cdots & g_{mn} \\
\vdots & \vdots    & \ddots & \vdots \\
g_{nm} & g_{n,m+1} & \cdots & g_{nn} 
\end{array}
\right)
\eea
vanishes. To prove this we shall display an elemnet
$\zeta' g \in H$. Since $h$ is a triangular matrix whose 
elements above the main diagonal are zero, $\zeta' g \in H$ if
\bea
\sum~ \zeta_{ps}'~g_{sq} ~=~ 0
\hspace{20mm} q>p
\eea
which implies
\bea
\tilde{G}_{p+1} ~ \zeta_p' ~=~ - \eta_p
\eea
where
\bea
\zeta_p'
~=~
\left(
\begin{array}{c}
\zeta_{p,p+1}' \\
\zeta_{p,p+2}' \\
\vdots \\
\zeta_{p,n}' 
\end{array}
\right)
\hspace{5mm}, \hspace{15mm}
\eta_p
~=~
\left(
\begin{array}{c}
g_{p,p+1} \\
g_{p,p+2} \\
\vdots \\
g_{p,n}
\end{array}
\right)
\eea
The existence of nontrivial solutions for $\zeta_{mn}'$,
therefore, requires that the determinant must be nonvanishing:
\bea
\mbox{det}~ \tilde{G}_{p+1}
~=~
\mbox{det}~ G_{p+1}
~=~
g_{p+1} ~\neq~ 0
\hspace{15mm}
p=1,2,3,\ldots
\eea
Hence if one (or more) of the minors $g_{p+1}$
vanishes the matrix $\tilde{G}_{p+1}$ ceases to be invertible
and, therefore, solutions for $\zeta'$ donot exist and the 
decomposition $g = \zeta h$ cannot exist.
Hence for $\zeta'$ thus determined
\bea
\zeta'~g &~=~& h \hspace{4mm}, \hspace{10mm} h \in H
\nonumber \\
g &~=~& \zeta'^{-1} h ~=~ \zeta ~ h
\eea
Analogously it may be shown that it is possible to write
\bea
g ~=~ k~x 
\hspace{4mm}, \hspace{10mm} k \in K
\hspace{3mm}, \hspace{3mm} x \in X
\eea
This decomposition exists if all the minors
\bea
g_m ~=~ \mbox{det}~ G_m ~\neq~ 0
\hspace{4mm}, \hspace{10mm} m=2,3,4,\ldots
\hspace{4mm}.
\eea
It now follows that multiplication of $g$ on the left by
$\zeta' \in Z$ keeps $\mbox{det}~ G_m = g_m$ invariant.
This can be written symbolically in the form
\bea
g_m (\zeta'~g) ~=~ g_m (g)
\eea
Writing $g = \zeta h$ we have
\bea
g_m (g) ~=~ g_m (\zeta'~ g) ~=~ g_m (\zeta'~\zeta~h)
\eea
where $\zeta'$ is completely arbitrary.
Hence choosing $\zeta' = \zeta^{-1}$ we have
\bea
g_m ( g) ~=~ g_m (h) ~=~ \mbox{det}~ H_m
\eea
where
\bea
H_m
&~=~&
\left(
\begin{array}{l@{~~~}l@{~~~}l@{~~~}c@{~~~}l}
h_{mm}    & 0           & 0         & \cdots & 0 \\
h_{m+1,m} & h_{m+1,m+1} & 0         & \cdots & 0 \\
\vdots    & \vdots      & \vdots    & \ddots & \vdots \\
h_{n,m}   & h_{n,m+1}   & h_{n,m+2} & \cdots & h_{nn} 
\end{array}
\right)
\eea
Hence
\bea
g_m ~=~ \mbox{det}~ H_m ~=~ h_{mm}~ h_{m+1,m+1}~\ldots ~h_{nn}
\eea
which yields
\bea
h_{mm} ~=~ \frac{g_m~~}{g_{m+1}}
\eea
Identical analysis can be carried out for the matrix $g x'$
and it now follows that 
\bea
k_{mm} ~=~ \frac{g_m~~}{g_{m+1}}
\eea
We have therefore shown for $g=kx$ and $g=\zeta h$
\bea
k_{pp} ~=~ h_{pp} ~=~ \frac{g_p~~}{g_{p+1}}
\label{eqn38}
\eea
Let us now denote by
\bea
\left(
\begin{array}{l@{~,~~}l@{~,~\ldots~,~~}l}
p_1 & p_2 & p_m \\
q_1 & q_2 & q_m 
\end{array}
\right)
\eea
the submatrix of $g$ consisting of the elements with row
indices $p_1, p_2, \ldots, p_m$ and column indices
$q_1, q_2, \ldots, q_m$ i.e.
\bea
\left(
\begin{array}{l@{~,~~}l@{~,~\ldots~,~~}l}
p_1 & p_2 & p_m \\
q_1 & q_2 & q_m
\end{array}
\right)
&=&
\left(
\begin{array}{c@{~~~}c@{~~~}c@{~~~}c}
g_{p_1,q_1} & g_{p_1,q_2} & \ldots & g_{p_1,q_m} \\
g_{p_2,q_1} & g_{p_2,q_2} & \ldots & g_{p_2,q_m} \\
\vdots      & \vdots      & \ddots & \vdots      \\
g_{p_m,q_1} & g_{p_m,q_2} & \ldots & g_{p_m,q_m} 
\end{array}
\right)
\eea
By repeating the previous arguments it can be shown that
\bea
\mbox{det}~
\left(
\begin{array}{l@{~,~~}l@{~,~~}l@{~,~\ldots~,~~}l}
p & p + 1 & p + 2 & n \\
q & p + 1 & p + 2 & n
\end{array}
\right)
&~=~&
h_{pq}~ h_{p+1, p+1}~ \ldots~ h_{nn}
\eea
Thus
\bea
h_{pq}
&~=~&
\mbox{det}~
\left(
\begin{array}{l@{~,~~}l@{~,~~}l@{~,~\ldots~,~~}l}
p & p + 1 & p + 2 & n \\
q & p + 1 & p + 2 & n
\end{array}
\right)
\Biggm/
g_{p+1}
\hspace{3mm}, \hspace{10mm} p>q
\eea
In a similar manner
\bea
k_{pq}
&~=~&
\mbox{det}~
\left(
\begin{array}{l@{~,~~}l@{~,~~}l@{~,~\ldots~,~~}l}
p & q + 1 & q + 2 & n \\
q & q + 1 & q + 2 & n
\end{array}
\right)
\Biggm/
g_{q+1}
\hspace{3mm}, \hspace{10mm} q>p
\eea
From the decomposition $k = \zeta \delta$ it now
follows that
\bea
\zeta_{pq}
&~=~&
\mbox{det}~
\left(
\begin{array}{l@{~,~~}l@{~,~~}l@{~,~\ldots~,~~}l}
p & q + 1 & q + 2 & n \\
q & q + 1 & q + 2 & n
\end{array}
\right)
\Biggm/
g_{q}
\hspace{3mm}, \hspace{10mm} q>p
\\[2mm]
x_{pq}
&~=~&
\mbox{det}~
\left(
\begin{array}{l@{~,~~}l@{~,~~}l@{~,~\ldots~,~~}l}
p & p + 1 & p + 2 & n \\
q & p + 1 & p + 2 & n
\end{array}
\right)
\Biggm/
g_{p}
\hspace{3mm}, \hspace{10mm} p>q
\label{eqn45}
\eea


\section{The Principal Series of Respresentations
of $SL(\lowercase{n},R)$}

The basic method of Gel'fand and coworkers for deriving
irreducible representations is the decomposition of a 
suitable reducible representation into irreducible parts.
We therefore introduce the reducible `quasiregular'
representation as follows:

Let us consider the Hilbert space $G(H)$ of square
integrable functions on $H$ with the scalar product
\bea
\left( f_1, f_2 \right)
&~=~&
\int
~\overline{f_1 (h)}
~f_2 (h) ~d \mu_r (h)
\eea
We first prove the following result.
The operator
\bea
T_g ~f(h) &~=~& f ( h_g)
\eea
is a unitary representation of $SL(n,R)$.
The matrix $h_g \in H$ is a matrix belonging to the coset
$\tilde{h}_g$.
By $\tilde{h}$ we mean the coset $Z_{g_o}$.
Under multiplication on the right by
$g \in SL(n,R)$ all elements of the coset
$\tilde{h}$ go into the elements of a single coset which
we denote by $\tilde{h}_g$. Thus $h' = h_g$ means that $h'$ and
$hg$ belong to the same coset i.e. we have
\bea
\zeta~ h' &~=~& h~g
\eea
It now easily follows that 
\bea
h_{g_1 g_2} &~=~& \left( h_{g_1} \right)_{g_2}
\eea
so that
\bea
T_{g_1 g_2} &~=~& T_{g_1} T_{g_2}
\eea
and $T_g$ is a representation.
We now show that $T_g$ is unitary i.e.
\bea
\left( T_g ~f_1, T_g ~f_2 \right)
&~=~&
\left( f_1, f_2 \right)
\eea
First we demonstrate that although the mapping 
$h \rightarrow h_g$ is not a right translation
the right invariant measure remains invariant  i.e.
\bea
d \mu_r (h)
&~=~&
d \mu_l (h')
\eea
To prove this we introduce the differential invariant
\bea
d \omega' &~=~& dh' \hspace{2mm} {h'\hspace{1mm}}^{-1}
\eea
\bea
d \mu_r (h') &~=~& \prod d \omega'_{pq}
\eea
We now arrange $d \omega'_{pq}$ in the following sequence
\bea
d\omega'_{21}, d\omega'_{31}, \ldots, d\omega'_{n1} ~;~
d\omega'_{22}, d\omega'_{32}, \ldots, d\omega'_{n2} ~;~
d\omega'_{n-1,n-1}, d\omega'_{n,n-1} ~;~ d\omega'_{nn}
\eea
Now since 
\bea
\zeta ~ h' &~=~& h~g
\eea
$\zeta$ and $h'$ are both functions of $h$.
We, therefore, obtain
\bea
\zeta~ dh' + d\zeta~ h'
&~=~&
dh~ g
\eea
This yields
\bea
dh' \hspace{2mm} {h'\hspace{1mm}}^{-1}
&~=~&
\zeta^{-1}~ dh~h^{-1} ~ \zeta
~-~
\zeta^{-1}~ d\zeta
\eea
so that
\bea
d \omega'
&~=~&
\zeta^{-1}~ d\omega~ \zeta
~-~
\zeta^{-1}~ d\zeta
~=~
- \zeta^{-1}~ d\zeta
~+~
du
\eea
where
\bea
du &~=~& \zeta^{-1}~ d\omega~ \zeta
\eea
Now we arrange both $du$ and $d\omega$ in the same
sequence as $d \omega'$ so that
\bea
\prod du_{pq}
&~=~&
\left| \mbox{det}~ D \right| ~\prod d \omega_{pq}
\eea
where $D$ is the Jacobian matrix
\bea
D &~=~&
\frac{
\partial ~(~ u_{21}, u_{31}, \ldots, u_{n1} ~;~
           u_{22}, u_{32}, \ldots, u_{n2} ~;~ \ldots ~;~
           u_{n-1,n-1}, u_{n,n-1} ~;~
           u_{nn} ~)
}{
\partial ~(~ \omega_{21}, \omega_{31}, \ldots, \omega_{n1} ~;~
           \omega_{22}, \omega_{32}, \ldots, \omega_{n2} ~;~ \ldots ~;~
           \omega_{n-1,n-1}, \omega_{n,n-1} ~;~
           \omega_{nn} ~)
}
\eea
It can be verified that $D$ is a block triangular matrix in which
each diagonal block is itself triangular and has the 
determinant $1$. Thus 
\bea
\mbox{det}~ D &~=~& 1
\eea
Hence
\bea
\prod_{p \geq q,~ p \neq 1} du_{pq}
&~=~&
d \mu_r (h)
\eea
Now 
\bea
d \omega'_{pq} &~=~&
du_{pq} ~-~
\sum_{r=p}^n~
\zeta'_{pr} ~ d\zeta_{rq}
\label{eqn56}
\eea
where we have written $\zeta^{-1} = \zeta'$.
For $p \geq q$ the second term of eqn. (\ref{eqn56})
is zero. Hence
\bea
d\omega'_{pq}
&~=~&
du_{pq}
\hspace{3mm}, \hspace{10mm} p \geq q
\eea
We therefore obtain
\bea
d \mu_r (h')
&~=~&
\prod_{p \geq q,~ p \neq 1} d\omega'_{pq}
~=~
\prod_{p \geq q,~ p \neq 1} du_{pq}
~=~
d \mu_r (h)
\eea
It now immediately follows that the quasiregular representation
is unitary
\bea
\left( T_g f_1, T_g f_2 \right)
&~=~&
\int d\mu_r (h)~ \overline{f_1 (h_g)} ~f_2(h_g)
\nonumber \\
&~=~&
\int d\mu_r (h')~ \overline{f_1 (h')} ~f_2(h')
\nonumber \\
&~=~&
\left( f_1, f_2 \right)
\eea
To decompose the quasiregular representation into 
irreducible representations we start from
\bea
\int f(h)~ d \mu_r (h)
&~=~&
\int \beta (h)~ f(h)~ d\mu_l (h)
\eea
If we introduce the decomposition
\bea
h &~=~& \delta~x
\eea
it then follows that
\bea
d \mu_l (h)
&~=~&
d \mu (\delta)~ d \mu (x)
\eea
Noting further that
\bea
\beta (\delta x) &~=~& \beta (\delta)~ \beta (x)
~=~ \beta (\delta)
\eea
we have
\bea
\int f(h)~ d \mu_r (h)
&~=~& 
\int d \mu (x)
\int f(\delta x)~ \beta (\delta)~ d\mu (\delta)
\eea
Replacing $f(h)$ by $\mid f(h) \mid^2$ we obtain
\bea
\int \mid f(h) \mid^2 d \mu_r (h)
&~=~&
\int d \mu (x)
\int \mid \phi (x, \delta) \mid^2 d \mu (\delta)
\eea
where
\bea
\phi (x, \delta)
&~=~&
f(\delta x)~ \beta^{\frac{1}{2}} (\delta)
\eea

Let $\sigma (\delta)$ be the character or equivalently
the one dimensional matrix element in a single irreducible
unitary representation of the commutative subgrup $D$.
We introduce the `Fourier transform' of the function
$\phi (x, \delta)$ in the following way:
\bea
f_{\sigma} (x)
&~=~&
\int \phi(x, \delta)~ \overline{\sigma} (\delta)~
d \mu (\delta)
\nonumber \\
&~=~&
\int \beta^{\frac{1}{2}} (\delta)~ f(\delta x)~
\overline{\sigma} (\delta)~ d \mu (\delta)
\eea
By Fourier transform we mean the $\sigma$~-~transform.
We shall presently see that `Fourier' transform in the 
context of real unimodular group is essentially
the Mellin transform. By Plancherel theorem it now
follows that
\bea
\int \mid f_{\sigma} (x) \mid^2 d \lambda (\sigma)
&~=~&
\int \mid \phi (x, \delta) \mid^2 d \mu (\delta)
\eea 
We, therefore, obtain
\bea
\parallel f \parallel^2
&~=~&
\int \mid f(h) \mid^2 d \mu_r (h)
\nonumber \\
&~=~&
\int d \mu (x)
\int \mid \phi (x, \delta) \mid^2 d \mu ( \delta)
\nonumber \\
&~=~&
\int d \lambda (\sigma) \int \mid f_{\sigma} (x) \mid^2 d \mu (x)
\eea
Thus 
\bea
\parallel f \parallel^2
&~=~&
\int d \lambda (\sigma)~ \parallel f_{\sigma} \parallel^2
\eea
This evidently means that $G(H)$ is decomposed into a
direct continuous sum of unitary spaces $G_{\sigma}(X)$.
We shall prove that this decomposition is simultaneously a
decomposition of the quasiregular representation $T_g$
into UIR's in each of the spaces $G_{\sigma}(X)$.

We denote the $\sigma$~-~component of the transformed function
$f'(h)= f(h_g)$ by $f'_{\sigma} (x)$
so that
\bea
f'(x)
&~=~&
\int f(h_g)~ \beta^{\frac{1}{2}} (\delta)~
\overline{\sigma (\delta)}~ d \mu (\delta)
\eea
where we have omitted the subscript $\sigma$.

If we now set $h_g = h^1$ it then follows that $hg$ and $h^1$
are in the same coset and we have
\bea
\zeta^1 ~h^1 &~=~& h~g
\eea
Introducing $h^1 = \delta^1 x^1$ we have
\bea
hg 
&~=~&
\delta^1~ \delta^1 \hspace{1mm}^{-1} ~\zeta^1 ~\delta^1 ~x^1
\eea
so that noting 
$\delta^1 \hspace{1mm}^{-1} \zeta^1 ~\delta^1 = \zeta ~\in ~Z$
we obtain
\bea
\delta ~ x ~ g &~=~& \delta^1 ~\zeta ~ x^1
\eea
where we have set $h=\delta x$. Hence
\bea
x ~ g
&~=~&
\delta^{-1} ~\delta^1 ~ \zeta~ x^1 ~=~ \delta^{(2)} ~\zeta ~ x^1
\eea
Since $\delta^{(2)} \zeta ~\in~K$
setting $k = \delta^{(2)} \zeta$
we have
\bea
x~g &~=~& k ~ x^1
\eea
This equation implies that $x^1$ and $xg$ lie in the same
right coset of $SL(n,R)$ by the subgroup $K$, i.e.
\bea
x^1 &~=~& x_g
\eea
Further
\bea
h_g &~=~& h^1 ~=~ \delta^1 ~ x^1 ~=~ \delta~ \delta^{(2)}~x_g
\eea
Hence we have
\bea
f'(x) &~=~&
\int f( \delta \delta^{(2)} x_g)~ \beta^{\frac{1}{2}} (\delta)
~\overline{\sigma (\delta)} ~d \mu (\delta)
\eea
Using the right invariance of the above integral and
the separability of the Radon~-~Nikodym derivative
\bea
f'(x)
&~=~&
\beta^{- \frac{1}{2}} (\delta^{(2)})~
\int f(\delta x_g)~ \beta^{\frac{1}{2}} (\delta)
~\overline{\sigma(\delta {\delta^{(2)}}^{-1})}
~d \mu (\delta)
\eea

Let us now consider some basic properties of the 
character $\sigma (\delta)$ of the group $D$.
Since the group $D$ is abelian its UIR's are one
dimensional and $\sigma (\delta)$ is the one dimensional 
matrix satisfying
\bea
\sigma (\delta \delta^{(2)})
&~=~&
\sigma (\delta)~ \sigma(\delta^{(2)})
\eea
which is the group composition law and
\bea
\overline{\sigma (\delta)} &~=~& \sigma(\delta^{-1})
\eea
which is the condition of unitarity of the representation.
These two equations imply that
\bea
\mid \sigma (\delta) \mid^2 &~=~& 1
\eea
Using these properties we obtain
\bea
\overline{\sigma(\delta {\delta^{(2)}}^{-1})}
&~=~&
\overline{\sigma} (\delta)
~\sigma ( \delta^{(2)} )
\eea
Thus 
\bea
f'(x)
&~=~&
\beta^{- \frac{1}{2}} (\delta^{(2)})
~\sigma ( \delta^{(2)} )
\int f(\delta x_g)~ \beta^{\frac{1}{2}} (\delta)
~\overline{\sigma(\delta )}
~d \mu (\delta)
\eea
the above integral is the $\sigma$~-~trasform of the 
function $f(\delta x_g)$. Thus 
\bea
T_g^{\sigma}~ f(x)
&~=~& 
\sigma ( \delta^{(2)} )
~\beta^{- \frac{1}{2}} (\delta^{(2)})
~f(x_g)
\eea
where  $\delta^{(2)}$ is defined by
\bea
x~g &~=~& \delta^{(2)} \zeta x^{1}
\hspace{3mm}, \hspace{10mm} k ~=~ \delta^{(2)} \zeta
\hspace{3mm}, \hspace{10mm} x^1 ~=~ x_g
\eea
It can be easily verfied that the operator 
$T_g^{\sigma}$ is unitary
\bea
(f_1, f_2)
&~=~&
(T_g^{\sigma} ~f_1, T_g^{\sigma} ~f_2)
\eea
To avoid inessential complication in the notation we shall
replace $\delta^{(2)}$ by $\delta$ so that
\bea
x~g &~=~& k x^1 ~=~ \delta ~\zeta ~x^1 
\hspace{3mm}, \hspace{10mm} x^1 ~=~ x_g 
\eea
The finite element of the group is then given by
\bea
T_g^{\sigma} ~f(x)
&~=~&
\sigma ( \delta )
~\beta^{- \frac{1}{2}} (\delta)
~f(x_g)
\eea
We now determine $\sigma ( \delta )$.
We note that the character $\sigma ( \delta )$ of the group $D$
is the character of the direct sum of
$n-1$ multiplicative groups of all real numbers
$\delta_2, \delta_3, \ldots, \delta_n$
and can be written as
\bea
\sigma (\delta)
&~=~&
\sigma_2 (\delta_2)
~\sigma_3 (\delta_3)
~\ldots
~\sigma_n (\delta_n)
\eea
Supressing the subscript we denote a particular factor
$\sigma_p (\delta_p)$ by $\sigma (\delta)$ which satisfies
\bea
\sigma (\delta)
&~=~&
\eta ( \mid \delta \mid ) ~\psi (\epsilon)
\eea
where
\bea
\delta &~=~& \mid \delta \mid ~\epsilon
\eea
and $\epsilon$ is the signature of $\delta$:
\bea
\epsilon &~=~& \frac{\delta}{\mid \delta \mid}
\eea
The function $\psi (\epsilon)$ must satisfy
\bea
\psi (\epsilon \epsilon') 
&~=~&
\psi (\epsilon) ~\psi (\epsilon')
\eea
Setting $\epsilon' = 1$ we have
\bea
\psi (\epsilon)
~\left[ \psi (1) - 1 \right]
=0
\eea
Hence $\psi (1)=1$.
Setting now $\epsilon = \epsilon'$ and noting that
$\epsilon^2 = 1$ we have
$\psi^2(\epsilon) = 1$.
Thus $\psi (\epsilon) = \pm 1$.
This evidently impies that
\bea
\psi (\epsilon) &~=~& 
\left( \frac{\delta}{\mid \delta \mid} \right)^{\eta}
\hspace{3mm}, \hspace{10mm} \eta = 0,1
\eea
To detrmine $\eta (\mid \delta \mid)$ we set 
$\mid \delta \mid = e^t$ so that
\bea
\eta (e^t) ~\eta (e^{t'})
 &~=~&
\eta (e^{t + t'})
\eea
which is solved by
\bea
\eta (e^t) = e^{i t \rho}
\hspace{10mm}
\mbox{i.e.}
\hspace{3mm}
\eta (\mid \delta \mid)
&~=~& 
\mid \delta \mid^{i \rho}
\eea
Combining these results we obtain
\bea
\sigma (\delta)
&~=~& 
\prod_{p=2}^{n} \sigma_p (\delta_p)
~=~ \prod_{p=2}^{n}
\mid \delta_p \mid^{i \rho_p}
\left( \frac{\delta_p}{\mid \delta_p \mid} \right)^{\eta_p}
\eea
Thus the finite element of the group is given by
\bea
T_g^{\sigma} ~f(x)
&~=~&
\mid \delta_2 \mid^{i \rho_2 - 1}
~\mid \delta_3 \mid^{i \rho_3 - 2}
~\ldots
~\mid \delta_n \mid^{i \rho_n - n + 1}
\nonumber \\
&& \hspace{20mm}
\left( \frac{\delta_2}{\mid \delta_2 \mid} \right)^{\eta_2}
~\left( \frac{\delta_3}{\mid \delta_3 \mid} \right)^{\eta_3}
~\ldots
~\left( \frac{\delta_n}{\mid \delta_n \mid} \right)^{\eta_n}
~f(x^1)
\eea
The above formula for the representation can be written out
in detail if in the groups $X$ and $SL(n,R)$ parameters
$x_{pq}$ $(p > q)$ and $g_{pq}$ are introduced.
In order to find $x^1 = x_g$ we must represent the element
\bea
g^1 &~=~& x~g
\eea
in the form
\bea
g^1 ~=~ x~g ~=~ k ~ x^1
\eea
Then $x^1 = x_g$. Using the eqn. (\ref{eqn45}) we have
\bea
x_{pq}^1
&~=~&
\mbox{det}~
\left(
\begin{array}{l@{~,~~}l@{~,~~}l@{~,~\ldots~,~~}l}
p & p + 1 & p + 2 & n \\
q & p + 1 & p + 2 & n
\end{array}
\right)^1
\Biggm/
g_{p}^1
\eea
In other words
\bea
x_{pq}^1
&~=~&
\mbox{det}~
\left(
\begin{array}{l@{~~~}l@{~~~}l@{~~~}l@{~~~}l}
g^1_{pq}    & g^1_{p,p+1}   & g^1_{p,p+2}   & \ldots & g^1_{pn}    \\
g^1_{p+1,q} & g^1_{p+1,p+1} & g^1_{p+1,p+2} & \ldots & g^1_{p+1,n} \\
\vdots      & \vdots        & \vdots        & \ddots & \vdots      \\
g^1_{n,q}   & g^1_{n,p+1}   & g^1_{n,p+2}   & \ldots & g^1_{n,n}   
\end{array}
\right)
\Biggm/
g_{p}^1
\eea
where
\bea
g^1_p
&~=~&
\mbox{det}~
\left(
\begin{array}{l@{~~~}l@{~~~}l@{~~~}l}
g^1_{pp}    & g^1_{p,p+1}   & \ldots & g^1_{pn}    \\
g^1_{p+1,p} & g^1_{p+1,p+1} & \ldots & g^1_{p+1,n} \\
\vdots      & \vdots        & \ddots & \vdots      \\
g^1_{n,p}   & g^1_{n,p+1}   & \ldots & g^1_{n,n}
\end{array}
\right)
\eea
In the above equations the elements $g_{pq}^1$ are given by,
\bea
g_{pq}^1 ~=~ g_{pq} ~+~ \sum_{r=1}^{p-1} ~x_{pr}~g_{rq}
\eea
It now remains to determine $\delta_p$, $p=2,3,\ldots,n$.
From eqns. (\ref{eqn22a}), (\ref{eqn22b}) and (\ref{eqn38}) 
\bea
\delta_p ~=~ k_{pp} ~=~ h_{pp} ~=~
\frac{g_p^1}{g_{p+1}^1}
\eea
The action of the finite element of the group is, therefore, given by
\bea
T_g^{\sigma}
~f(x)
&~=~&
\mid \frac{g_2^1}{g_3^1} \mid^{i \rho_2 - 1}
~\mid \frac{g_3^1}{g_4^1} \mid^{i \rho_3 - 2}
~\ldots
~\mid \frac{g_{n-1}^1}{g_n^1} \mid^{i \rho_{n-1}- n+2}
~\mid g_n^1 \mid^{i \rho_{n}- n+1}
\nonumber \\
&& \hspace{20mm}
\mbox{sgn}^{\eta_2} \left( \frac{g_2^1}{g_3^1} \right)
~\mbox{sgn}^{\eta_3} \left( \frac{g_3^1}{g_4^1} \right)
~\ldots
~\mbox{sgn}^{\eta_n} \left( g_n^1 \right)
~f(x^1)
\eea
 

\section{The character of the principal series of 
representations of $SL(\lowercase{n},R)$} 

We now proceed to compute the character of the above representations.
We introduce the operator of the group ring
\bea
T_t^{\sigma} ~=~
\int d \mu (g) ~ t(g) ~T_g^{\sigma}
\eea
where $t(g)$ is an arbitrary test function on the group having
a compact support. The action of the group ring
is then given by
\bea
T_t^{\sigma} ~f(x) 
&~=~&
\int d \mu (g) ~ t(g) 
~\beta^{- \frac{1}{2}} (\delta)
~\sigma (\delta)
~f (x^1)
\eea
Let us now perform the left translation 
$g \rightarrow x^{-1} g$.
Under this mapping the equation $x g = k x^1$
is replaced by $g = k x_1$ where
\bea
{(x_1)}_{pq}
&~=~&
\mbox{det}~
\left(
\begin{array}{l@{~~~}l@{~~~}l@{~~~}l}
g_{pq}    & g_{p,p+1}   & \ldots & g_{pn}    \\
g_{p+1,q} & g_{p+1,p+1} & \ldots & g_{p+1,n} \\
\vdots    & \vdots      & \ddots & \vdots    \\
g_{nq}   & g_{n,p+1}   & \ldots & g^1_{nn}
\end{array}
\right)
\Biggm/
g_{p}
\eea
It can be shown that under the decomposition $g = k x_1$
the invariant measure decomposes as
\bea
d \mu_l(g) ~=~ d \mu_r(g) 
~=~
d \mu_l(k) ~d \mu (x_1) 
\eea
Thus 
\bea
T_t^{\sigma} ~f(x)
&~=~&
\int d \mu (x_1)
\int d \mu_l (k)
~t(x^{-1} k x_1)
~ \mid k_{22} \mid^{i \rho_2 -1}
~ \mid k_{33} \mid^{i \rho_3 -2}
~\ldots
~ \mid k_{nn} \mid^{i \rho_n -n+1}
\hspace{10mm}
\nonumber \\
&& \hspace{50mm}
~ \mbox{sgn}^{\eta_2} ~k_{22}
~ \mbox{sgn}^{\eta_3} ~k_{33}
~\ldots
~ \mbox{sgn}^{\eta_n} ~k_{nn}
~f(x_1)
\eea
which can be written in the form
\bea
T_t^{\sigma} ~f(x)
&~=~&
\int 
K(x, x_1)
~f(x_1)
~d \mu (x_1)
\eea
The integral kernel of the group ring is,
therefore, given by
\bea
K(x,x_1)
&~=~&
\int_K ~t(x^{-1} k x_1)
~ \mid k_{22} \mid^{i \rho_2 -1}
~ \mid k_{33} \mid^{i \rho_3 -2}
~\ldots
~ \mid k_{nn} \mid^{i \rho_n -n+1}
\hspace{10mm}
\nonumber \\
&& \hspace{50mm}
~ \mbox{sgn}^{\eta_2} ~k_{22}
~ \mbox{sgn}^{\eta_3} ~k_{33}
~\ldots
~ \mbox{sgn}^{\eta_n} ~k_{nn}
~d \mu_l (k)
\label{eqn113}
\eea
The above kernel has a trace $\mbox{Tr}~ (T_t^{\sigma})$
given by
\bea
\mbox{Tr}~ (T_t^{\sigma})
&~=~&
\int K(x,x) ~d \mu (x)
\eea
The eqn. (\ref{eqn113}), therefore yields
\bea
\mbox{Tr}~ (T_t^{\sigma})
&~=~&
\int_K \int_X
~t(x^{-1} k x)
~ \mid k_{22} \mid^{i \rho_2 -1}
~ \mid k_{33} \mid^{i \rho_3 -2}
~\ldots
~ \mid k_{nn} \mid^{i \rho_n -n+1}
\hspace{10mm}
\nonumber \\
&& \hspace{40mm}
~ \mbox{sgn}^{\eta_2} ~k_{22}
~ \mbox{sgn}^{\eta_3} ~k_{33}
~\ldots
~ \mbox{sgn}^{\eta_n} ~k_{nn}
~d \mu_l (k)
~d \mu (x)
\eea
We now note that the elements of the group $SL(n,R)$
with distinct eigenvalues can be divided into two broad
classes:
\begin{enumerate}
\item[(a)]
the `hyperbolic' class for which the eigenvalues
of the matrix $g$ are real,
\item[(b)]
the `elliptic' class for which the eigenvalues of $g$
are complex.
\end{enumerate}

We shall first show that every hyperbolic element
of $SL(n,R)$ can be represented in the form
\bea
g ~=~ x^{-1} ~k ~x
\eea 
where $k_{pp} = \lambda_p$, ($p=1, 2, 3, \ldots, n$)
are the real eigenvalues of the matrix $g$ taken in any order.

We recall that every $g \in SL(n,R)$ belonging to the 
hyperbolic class can be diagonalized as 
\bea
g' ~ g ~{g'\hspace{1mm}}^{-1} ~=~ \delta
\eea
where
\bea
\delta 
&~=~&
\left(
\begin{array}{c@{~~~}c@{~~~}c@{~~~}c@{~~~}c}
\delta_1 & 0       & 0      & \cdots & 0 \\
0        & \delta_2 & 0      & \cdots & 0 \\
\vdots   & \vdots  & \vdots & \ddots & \vdots \\
0        & 0       & 0      & \cdots & \delta_n
\end{array}
\right)
\eea
belongs to the subgroup of real diagonal matrices of
determinant unity and 
$g' \in SL(n,R)$. We now use the decomposition
\bea
g' ~=~ k' ~ x
\eea
Thus
\bea
g ~=~ {g'\hspace{1mm}}^{-1} ~\delta ~g'
~=~ x^{-1} ~ {k'\hspace{1mm}}^{-1} ~\delta ~k' ~x 
\eea
Now ${k'\hspace{1mm}}^{-1} ~\delta ~k' ~\in ~K$,
so that writing 
$k = {k'\hspace{1mm}}^{-1} ~\delta ~k'$
we have the decomposition 
$g=x^{-1} k x$
in which $k_{pp} = \delta_p = \lambda_p$,
$p=1, 2, \ldots, n$.
We can, therefore, say that every hyperbolic element
$g \in SL(n,R)$ can be represented in the form
\bea
g = x^{-1} ~ k ~x
\eea
where
\bea
k_{pp} ~=~ \lambda_g^{(p)}
\hspace{3mm}, \hspace{10mm}
p = 1, 2, 3, \ldots, n.
\eea
are the real eigenvalues of the matrix $g$ taken in any order.
We can therefore assert that for the principal series
of representations the trace is concentrated on the hyperbolic
elements.

We now proceed to derive the integral relation
connected with the representation of g in the form
$g = x^{-1} k x$ and show that
\bea
\mbox{Tr}~ (T_t^{\sigma})
&~=~&
\int_X d \mu (x)
\int_K d \mu_l (k)
~t(x^{-1} k x) ~\phi_{\rho} (k)
\nonumber \\
&~=~&
\int t(g)
~\frac{\sum ~\phi_{\rho} (k_g) ~\beta^{\frac{1}{2}} (k_g)}
{\prod_{p>q} \mid \lambda_g^{(p)} - \lambda_g^{(q)} \mid}
~d \mu (g)
\label{eqn122a}
\eea
where
\bea
\phi_{\rho} (k)
&~=~&
\mid k_{22} \mid^{i \rho_2 -1}
~ \mid k_{33} \mid^{i \rho_3 -2}
\ldots
\mid k_{nn} \mid^{i \rho_n -n+1}
~ \mbox{sgn}^{\eta_2} ~k_{22}
~ \mbox{sgn}^{\eta_3} ~k_{33}
~\ldots
~ \mbox{sgn}^{\eta_n} ~k_{nn}
,
\label{eqn122b}
\eea
$t(g)$ is a function on $SL(n,R)$,
$k_g$ are the elements of $k$ such that
$x^{-1} k_g x = g$,
$\beta (k) = \frac{d \mu_l (k)}{d \mu_r (k)}$
and the sum is taken over all $k_g$ which are
derived by all possible permutations on the main
diagonal of $k_g$ of the eigenvalues
of the element $g$; finally
$\lambda_g^{(1)}, \lambda_g^{(2)}, \ldots, \lambda_g^{(n)}$
are the eigenvalues of $g$.

To prove the above formula we remove from the group $K$
all those matrices $k$ for which the moduli of any two
eigenvalues coincide.
This at the same time cuts $K$ into $n!$ connected 
(but dosjoint) regions $K_s$ such that each of these regions contains
no pair of matrices $k$, the diagonal elements of which 
differ only in order. Since in cutting $K$
into $K_s$ only manifolds of lower dimension are removed,
the integral over $K$ is decomposed into the sum of 
$n!$ integrals over $K_s$.

We now set $g= x^{-1} k x$, $x \in X$, $k \in K_s$.
If $k$ runs over $K_s$ and $x$ runs over $X$ the element
$g$ runs once over all the hyperbolic elements of 
$SL(n,R)$ except those for which one of the minors
$x_m$ equals zero.
But these excluded elements make up a manifold of lower dimensions;
consequently elements of the form $g = x^{-1} k x$
fill up all the hyperbolic elements of $SL(n,R)$ except
of a set of lower dimensions.
We now find the relation between the invariant measures
in $SL(n,R)$, $K$ and $X$ under the condition
$g = x^{-1} k x$.

We first introduce the differential invariant
\bea
d \omega^g ~=~ g^{-1}~ dg ~=~ x^{-1} ~ du ~ x 
\label{eqn123}
\eea
where
\bea
du ~=~ d \omega^k ~+~ d \omega^x ~-~ k^{-1}~ d \omega^x ~k
\eea
\bea
d \omega^k ~=~ k^{-1} ~dk
\hspace{3mm}, \hspace{15mm}
d \omega^x ~=~ dx~ x^{-1}
\label{eqn125}
\eea
we shall first prove 
\bea
d \mu (g) ~=~ \prod d \omega_{pq}^g
~=~ \prod du_{pq}
\eea
where the product is taken over all the independent
elements of $d \omega^g$ and $du$.

To prove the above formula we arrange the components of 
$d \omega^g$ in the following order,
\bea
d \omega_{12}^g, d \omega_{13}^g, \ldots, d \omega_{1n}^g ~;~ 
d \omega_{22}^g, d \omega_{23}^g, \ldots, d \omega_{2n}^g ~;~ 
\ldots ~;~
d \omega_{n-1,n-1}^g, d \omega_{n-1,n}^g ~;~ 
d \omega_{n,n}^g 
\hspace{10mm}
\nonumber \\
d \omega_{n1}^g, d \omega_{n-1,1}^g, \ldots, d \omega_{21}^g ~;~ 
d \omega_{n2}^g, d \omega_{n-1,2}^g, \ldots, d \omega_{32}^g ~;~ 
\ldots ~;~
d \omega_{n,n-1}^g, d \omega_{n-1,n-1}^g ~;~ 
d \omega_{n,n-1}^g 
\nonumber
\eea
and also the components of $du$ in the same order. Thus
\bea
d \mu (g) ~=~ \prod ~ d \omega^g
~=~ \mid \mbox{det}~ D \mid
~\prod ~ du
\eea 
Here $D$ is a block triangular matrix in which each diagonal
block except the last is itself triangular and has the
determinant $1$. Thus calling the last block $L$ we have
\bea
\mbox{det}~ D ~=~ \mbox{det}~ L
\eea
where $L$ equals to the matrix
\bea
\left(
\begin{array}{llclllclllcl}
x'_{n1} \frac{\partial u_{11}}{\partial u_{n1}} +1 &
x'_{n-1,1} \frac{\partial u_{11}}{\partial u_{n1}} &
\dots &
x'_{21} \frac{\partial u_{11}}{\partial u_{n1}} &
0 & 0 & \ldots & 0 & 0& 0 & \ldots & 0 \\
x'_{n1} \frac{\partial u_{11}}{\partial u_{n-1,1}} + x'_{n,n-1} &
x'_{n-1,1} \frac{\partial u_{11}}{\partial u_{n-1,1}}+ 1 &
\dots &
x'_{21} \frac{\partial u_{11}}{\partial u_{n-1,1}} &
0 & 0 & \ldots & 0 & 0& 0 & \ldots & 0 \\
\hspace{10mm} \vdots & \hspace{10mm} \vdots & \ddots 
& \hspace{7mm} \vdots & \vdots & \vdots &
\ddots & \vdots & \vdots & \vdots & \ddots & \vdots \\ 
x'_{n1} \frac{\partial u_{11}}{\partial u_{21}} + x'_{n2} &
x'_{n-1,1} \frac{\partial u_{11}}{\partial u_{21}}+ x'_{n-1,2} &
\dots &
x'_{21} \frac{\partial u_{11}}{\partial u_{21}} + 1 &
0 & 0 & \ldots & 0 & 0& 0 & \ldots & 0 \\
x'_{n1} \frac{\partial u_{11}}{\partial u_{n2}} + x_{21} &
x'_{n-1,1} \frac{\partial u_{11}}{\partial u_{n2}} &
\dots &
x'_{21} \frac{\partial u_{11}}{\partial u_{n2}} &
1 & 0 & \ldots & 0 & 0& 0 & \ldots & 0 \\

x'_{n1} \frac{\partial u_{11}}{\partial u_{n-1,2}} + x'_{n,n-1} x_{21} &
x'_{n-1,1} \frac{\partial u_{11}}{\partial u_{n-1,2}} + x_{21}&
\dots &
x'_{21} \frac{\partial u_{11}}{\partial u_{n-1,2}} &
x'_{n,n-1} & 1 & \ldots & 0 & 0& 0 & \ldots & 0 \\
\hspace{10mm} \vdots & \hspace{10mm} \vdots & \ddots  
& \hspace{7mm} \vdots & \vdots & \vdots &
\ddots & \vdots & \vdots & \vdots & \ddots & \vdots \\
x'_{n1} \frac{\partial u_{11}}{\partial u_{32}} + x'_{n3} x_{21} &
x'_{n-1,1} \frac{\partial u_{11}}{\partial u_{32}} +x'_{n-1,3} x_{21} &
\dots &
x'_{21} \frac{\partial u_{11}}{\partial u_{32}} &
x'_{n3} & x'_{n-1,3} & \ldots & 1 & 0& 0 & \ldots & 0 \\
\hspace{10mm} \vdots & \hspace{10mm} \vdots & \ddots
& \hspace{7mm} \vdots & \vdots & \vdots &
\ddots & \vdots & \vdots & \vdots & \ddots & \vdots \\
x'_{n1} \frac{\partial u_{11}}{\partial u_{n,n-1}} + x_{n-1,1} &
x'_{n-1,1} \frac{\partial u_{11}}{\partial u_{n,n-1}} &
\dots &
x'_{21} \frac{\partial u_{11}}{\partial u_{n,n-1}} &
x'_{n-1,2} & 0 & \ldots & 0 & 0& 0 & \ldots & 1 
\end{array}
\right)
\label{eqn128}
\eea
where we have written 
\bea
x^{-1} ~=~ x'
\eea
We shall show that
\bea
\frac{\partial u_{11}}{\partial u_{p1}} ~=~ 0
\hspace{3mm}, \hspace{10mm}
p = 2, 3, \ldots, n
\eea
so that $L$ becomes a traingular matrix with detrminant $1$.
To prove this we start from 
\bea
d \omega_{pq}^g ~=~
\sum_{r=1}^n ~ (g^{-1})_{pr} ~ dg_{rq}
~=~ \sum_{r=1}^n ~ a_{rp} ~ dg_{rq} 
\eea
where $a_{rp}$ is the cofactor of $g_{rp}$.
Setting $p=q=1$ we have 
\bea
d \omega_{11}^g ~=~
a_{11} ~ dg_{11}
~+~ \sum_{r=2}^n ~ a_{r1} ~ dg_{r1}
\eea
In the above it should be noted that although $dg_{11}$
is not independent all other $dg_{rp}$ are independent
differentials. Hence
\bea
\frac{\partial \omega_{11}^g}{\partial g_{p1}}
~=~
a_{11} ~ \frac{\partial g_{11}}{\partial g_{p1}}
~+~ a_{p1} 
\label{eqn132}
\eea
To calculate $\frac{\partial g_{11}}{\partial g_{p1}}$
we expand $\mbox{det}~g$ ($=1$) with respect to the 
first column so that
\bea
g_{11}~ a_{11} ~+~
g_{21}~ a_{21} ~+~
g_{31}~ a_{31} ~+~
\ldots ~+~
g_{n1}~ a_{n1} 
~=~ 1
\eea
We now note that none of he cofactors appearing above 
contains elements from the first column i.e. all the 
cofactors are independent of 
$g_{21}, g_{31}, g_{41}, \dots, g_{n1}$.
Hence differentiating with respect to $g_{p1}$ we have
\bea
\frac{\partial g_{11}}{\partial g_{p1}}
~a_{11} ~+~ a_{p1} ~=~ 0
\eea
Thus 
\bea
\frac{\partial g_{11}}{\partial g_{p1}}
&~=~&
- ~\frac{a_{p1}}{a_{11}}
\label{eqn135}
\eea
Substituting eqn. (\ref{eqn135}) in eqn. (\ref{eqn132}) we have
\bea
\frac{\partial \omega_{11}^g}{\partial g_{p1}}
~=~
a_{11} ~ \frac{\partial g_{11}}{\partial g_{p1}}
~+~ a_{p1}
~=~ 0
\label{eqn136}
\eea
Let us now consider 
\bea
d \omega_{1q}^g
&~=~&
\sum ~ (g^{-1})_{1r} ~dg_{rq}
~=~ a_{11} ~dg_{1q} ~+~ 
a_{21} ~dg_{2q} ~+~
\ldots ~+~
a_{n1} ~dg_{nq} 
\eea
Thus 
\bea
\frac{\partial \omega_{1q}^g}{\partial g_{p1}}
~=~ 0
\hspace{3mm}, \hspace{10mm}
q = 2, 3, 4, \ldots, n 
\hspace{3mm}, \hspace{10mm}
p = 2, 3, 4, \ldots, n 
\label{eqn138}
\eea
Now we write eqn. (\ref{eqn123}) in the form 
$du = x ~d\omega^g ~x^{-1}$ so that 
\bea
du_{11} ~=~
d\omega^g_{11} ~+~
\sum_{q=2}^n ~d\omega^g_{1q} ~x'_{q1}
\hspace{3mm}, \hspace{10mm}
x' ~=~ x^{-1}
\eea
Thus 
\bea
\frac{\partial u_{11}}{\partial g_{p1}}
~=~
\frac{\partial \omega_{11}^g}{\partial g_{p1}}
~+~
\sum_{q=2}^n
~ \frac{\partial \omega_{1q}^g}{\partial g_{p1}}
~ x'_{q1}
~=~ 0
\hspace{3mm}, \hspace{10mm}
p = 2, 3, 4, \ldots, n 
\label{eqn140}
\eea
where we have used eqn. (\ref{eqn136}) and eqn. (\ref{eqn138}).
We now note
\bea
\frac{\partial u_{11}}{\partial u_{p1}}
~=~
\sum_{k=2}^n
~\frac{\partial u_{11}}{\partial g_{k1}}
~\frac{\partial g_{k1}}{\partial u_{p1}}
~+~
\sum_{k=1}^n \sum_{l=2}^n
~\frac{\partial u_{11}}{\partial g_{kl}}
~\frac{\partial g_{kl}}{\partial u_{p1}}
\label{eqn141}
\eea
The eqn. (\ref{eqn140}) implies that the first term on the 
r. h. s. of eqn. (\ref{eqn141}) is zero so that
\bea
\frac{\partial u_{11}}{\partial u_{p1}}
~=~
\sum_{k=1}^n \sum_{l=2}^n
~\frac{\partial u_{11}}{\partial g_{kl}}
~\frac{\partial g_{kl}}{\partial u_{p1}}
\eea
We shall now show that
\bea
\frac{\partial g_{kl}}{\partial u_{p1}}
~=~ 0
\hspace{10mm} \mbox{for} \hspace{10mm}
k = 1, 2, 3, \ldots, n
\hspace{3mm}, \hspace{10mm}
l = 2, 3, \ldots, n
\label{eqn142}
\eea
To prove this we write eqn. (\ref{eqn123}) in the form
\bea
dg ~=~ g ~x^{-1} ~du ~x ~=~ g' ~ du ~x
\eea
where we have written $g x^{-1} = g'$. Thus
\bea
dg_{kl}
&~=~&
\sum_{r=1}^n \sum_{s=l}^n
~g'_{kr} ~du_{rs}  ~x_{sl}
\eea
so that 
\bea
\frac{\partial g_{kl}}{\partial u_{p1}}
~=~
\sum_{r=1}^n \sum_{s=l}^n
~g'_{kr}
~\frac{\partial u_{rs}}{\partial u_{p1}}
~x_{sl}
\eea
Now, in the sum on the r. h. s. of eqn. (\ref{eqn142})
$l$ starts from $2$. Hence for 
$l = 2, 3, 4, \ldots, n$
\bea
\frac{\partial g_{kl}}{\partial u_{p1}}
~=~
\sum_{r=1}^n
~g'_{kr}
\sum_{s=l}^n
~\frac{\partial u_{rs}}{\partial u_{p1}}
~x_{sl}
~=~ 0
\eea
because
\bea
\frac{\partial u_{rs}}{\partial u_{p1}} ~=~ 0
\hspace{10mm} \mbox{for} \hspace{10mm}
s = 2, 3, 4, \ldots, n
\eea
Hence finally we have
\bea
\frac{\partial u_{11}}{\partial u_{p1}} ~=~ 0
\hspace{10mm} \mbox{for} \hspace{10mm}
p = 2, 3, 4, \ldots, n
\eea
and the matrix $L$ in eqn. (\ref{eqn128}) becomes a 
triangular matrix with $1$ along tha main diagonal so that 
$\mbox{det}~L = 1$ and we have
\bea
d \mu (g) ~=~ \prod ~d \omega^g ~=~ \prod  ~du
\eea
Here
\bea
du ~=~ d \omega^k ~+~ d \omega^x
~-~ k^{-1} ~d \omega^x ~k
~=~ d \omega^k ~+~ dv
\label{eqn149}
\eea
where
\bea
dv ~=~ d \omega^x
~-~ k^{-1} ~d \omega^x ~k
\eea
and $d \omega^k$ and $d \omega^x$ are given by
eqn. (\ref{eqn125}).

In eqn. (\ref{eqn149}) $d \omega^k$ is a triangular matrix with 
$k_{11}^{-1} dk_{11}, k_{22}^{-1} dk_{22}, \ldots, k_{nn}^{-1} dk_{nn}$
along the principal diagoanl.
$dv$ is a square matrix in which the independent elements are
\bea
dv_{21} ~;~ 
dv_{31}, dv_{32} ~;~
dv_{41}, dv_{42}, dv_{43} ~;~
\ldots ~;~
dv_{n1}, dv_{n2}, \dots, dv_{n,n-1}.
\eea
The elements like 
$dv_{12}, dv_{13}, \ldots$ are nonzero but not 
independent and for all the elements
\bea
\frac{\partial v_{pq}}{\partial \omega^k_{rs}} ~=~ 0
\eea
Thus 
\bea
d \mu (g)
~=~ \prod ~du
~=~ \mid \mbox{det}~ D \mid ~\prod ~d \omega^k ~\prod ~dv
\label{eqn152}
\eea
where $D$ is the Jacobian matrix.
If we now note that
\bea
du_{pq} &~=~& d \omega^k_{pq} + dv_{pq}
\hspace{10mm} p\leq q
\\
du_{pq} &~=~& d v_{pq}
\hspace{23mm} p > q
\eea
the matrix $D$ in eqn. (\ref{eqn152}) becomes a traingular 
matrix with $1$ along the main diagonal so that
\bea
d \mu (g) ~=~ d \mu_l (k) ~\prod_{p>q} ~dv_{pq}
\eea
where 
\bea
dv ~=~ d \omega^x ~-~ k^{-1} ~d \omega^x ~k
\eea
Let us now set
\bea
k ~=~ \zeta^{-1} ~ \delta ~\zeta
\eea
We then obtain
\bea
dv ~=~ \zeta^{-1} ~ dp ~\zeta
\label{eqn156}
\eea
where
\bea
dp &~=~& d w ~-~ \delta^{-1} ~d w ~\delta
\\
d w &~=~& \zeta ~ d \omega^x \zeta^{-1}
\eea
Writing $\zeta^{-1} = \zeta'$ we obtain from eqn. (\ref{eqn156})
\bea
dv_{pq}
~=~
\sum_{r=p}^n
\sum_{s=1}^q
~\zeta'_{pr} ~ dp_{rs} ~ \zeta_{sq}
\hspace{20mm} p>q
\eea
in which $r_{min} = p$, 
$s_{max} = q$,
$r_{min} > s_{max}$, so that
\bea
\frac{\partial v_{pq}}{\partial p_{ml}}
~=~ \zeta'_{pm} ~\zeta_{lq}
\eea
It, therefore, follows that the determinant of the matrix
connecting $dv_{pq}$ ($p>q$) and 
$dp_{ml}$ ($m>l$) is a block triangular determinant in which
each diagonal block itself is triangular and has the 
determinant $1$. Thus
\bea
\prod_{r>s} ~dv_{rs} ~=~ \prod_{r>s} ~dp_{rs}
\eea
so that 
\bea
d \mu (g) ~=~ d \mu_l (k) ~ \prod_{r>s} ~dp_{rs}
\eea
Now
\bea
dp_{rs} ~=~ d w_{rs}
\left( 1 - \frac{\delta_s}{\delta_r} \right)
\eea
and 
\bea
\prod_{r>s} ~dp_{rs} ~=~
\mid \mbox{det}~ D \mid ~\prod_{r>s} ~d w_{rs}
\eea
where $D$ is a diagonal matrix and $\mbox{det}~ D$ is the 
product of its diagonal elements. Thus
\bea
\mbox{det}~ D
&~=~&
\prod_{p>q} ~(\delta_p - \delta_q)
~\left[
\delta_2 ~\delta_3^2 ~\delta_4^3 \ldots \delta_n^{n-1}
\right]^{-1}
\eea
Since $\delta_p = k_{pp}$ the above equaton yields
\bea
\prod_{r>s} ~dp_{rs} ~=~
\prod_{p>q} 
~\frac{\mid k_{pp} - k_{qq} \mid}
	{\mid k_{22} \mid \mid k_{33} \mid^2 \ldots \mid k_{nn} \mid^{n-1}}
~\prod_{r>s} ~d w_{rs}
\eea
Since the denominator is the square root of the 
Radon~-~Nikodym derivative we can write
\bea
\prod_{r>s} ~dp_{rs} ~=~
\prod_{p>q}
~\frac{\mid k_{pp} - k_{qq} \mid}
        {\beta^{\frac{1}{2}} (k)}
~\prod_{r>s} ~d w_{rs}
\eea
It is now easy to check that the transformation from 
$d \omega^x$ to 
$d w = \zeta ~d \omega^x ~\zeta^{-1}$ is a linear mapping of the 
components of $d \omega^x$ with determinant $1$. Thus
\bea
\prod_{r>s} ~d w_{rs} ~=~
\prod_{r>s} ~d \omega_{rs}^x ~=~
d \mu (x)
\eea
We, therefore finally obtain
\bea
d \mu (g)
~=~
\prod_{p>q} 
~\frac{\mid k_{pp} - k_{qq} \mid}
        {\beta^{\frac{1}{2}} (k)}
~d \mu_l (k) ~d \mu (x)
\eea
so that
\bea
d \mu_l (k) ~d \mu (x)
~=~
\frac{\beta^{\frac{1}{2}} (k_g)}
	{\prod_{p>q} \mid \lambda_g^{(p)} - \lambda_g^{(q)} \mid}
~d \mu (g)
\eea
Hence finally 
\bea
\int_X d \mu (x)
\int_{K_s} t(x^{-1} k x)
~\phi_{\rho} (k) ~d \mu_l (k)
&~=~&
\int t(g) ~\phi_{\rho} (k_g)
~\frac{\beta^{\frac{1}{2}} (k_g)}
        {\prod_{p>q} \mid \lambda_g^{(p)} - \lambda_g^{(q)} \mid}
~d \mu (g)
\eea
where $\phi_{\rho} (k_g)$ and 
$\beta^{\frac{1}{2}} (k_g)$ are given by
eqns. (\ref{eqn122b}) and (\ref{eqn13}) respectively and
$k_g$ is taken from $K_s$. Summing this equation over all $K_s$
we obtain the formula in eqn. (\ref{eqn122a}).
We can now immediately write down the character $\pi (g)$ of the
representation as
\bea
\pi (g)
~=~
\frac{\sum \chi_{\rho} (k_g)}
	{\prod_{p>q} \mid \lambda_g^{(p)} - \lambda_g^{(q)} \mid}
\label{eqn171}
\eea
where
\bea
\chi_{\rho} (k_g)
~=~
\prod_{p=2}^n ~\mid \lambda_g^{(p)} \mid^{i \rho_p}
~sgn^{\eta_p} ~\lambda^p_g
\eea
It can be easily verified that the formula (\ref{eqn171})
yields the formula  for the character of the 
principal series of representations of $SL(2,R)$ for
$n=2$.
\bea
\pi (g) ~=~ 
\frac{\mid \lambda_g \mid^{i \rho} + \mid \lambda_g \mid^{-i \rho}} 
	{\mid \lambda_g - \lambda_g^{-1} \mid}
~sgn^{\eta} ~\lambda_g
\eea
in agreement with ref. \cite{basu_1996}.



\vspace{-135mm}
\begin{center}
{\large \bf References}
\end{center}



\begin{references}

\newpage
\vspace*{10mm}

\bibitem{gelf_graev}
I. M. Gel'fand and M. I. Graev,
{\it I. M. Gel'fand~-~Collected Papers}, 
Springer~-~Verlag, Berlin, 1988. 
\bibitem{gelf_naimark}
I. M. Gel'fand and M. A. Naimark, 
{\it Unitary Representations of Semisimple Lie Groups I},
American Mathematical Society, 
Translation Number 64, 1952.
\bibitem{gelf_naimark_paper}
I. M. Gel'fand and M. A. Naimark,
ref. \cite{gelf_graev}, p41,182.
\bibitem{basu_1996}
S. Bal, K. V. Shajesh and D. Basu,
J. Math. Phys. {\bf 38}, 3209 (1997).
\bibitem{basu_1999}
D. Basu, S. Bal and K. V. Shajesh,
J. Math. Phys. {\bf 41}, 461 (2000).
\bibitem{bargmann199a}
V. Bargmann, 
Commun. Pure Appl. Math. {\bf 14}, 187 (1961);
{\bf 20}, 1 (1967); 
in {\it Analytic Methods in Mathematical Physics},
edited by P. Gilbert and R. G. Newton.
\bibitem{segal}
I. E. Segal,
Ill. J. Math. {\bf 6}, 500 (1962).
\bibitem{bargmann199b}
V. Bargmann,
Ann. Math. {\bf 48}, 568 (1947).
\end{references}
\end{document}